\documentclass[11pt,a4paper]{article}
\usepackage[T1]{fontenc}
\usepackage{lmodern}
\usepackage[margin=25mm]{geometry}
\usepackage{amsmath,amssymb,booktabs,graphicx,tabularx,array}
\usepackage[unicode,hidelinks]{hyperref}
\usepackage{microtype}
\title{Post-processing probabilities of exceeding historical heat records}
\author{Yucan Wang\thanks{Corresponding author. Email: \href{mailto:wyclemon@foxmail.com}{wyclemon@foxmail.com}}\\
Li Zhang\thanks{Email: \href{mailto:421138336@qq.com}{421138336@qq.com}}\\
Yang Hou\thanks{Email: \href{mailto:19953223040@163.com}{19953223040@163.com}}\\[4pt]
\small Henan Province Meteorological Data Center, 110 Jinshui Road,\\
Jinshui District, Zhengzhou, Henan Province, China}
\date{}
\begin{document}
\maketitle

\begin{abstract}
Statistical post-processing can improve the overall temperature distribution while distorting probabilities in its upper tail. We verify 2022 medium-range forecasts of daily maximum temperature over land against local monthly maxima from ERA5 in 1979--2017. At days 1--7, rolling Gaussian ensemble model output statistics (EMOS) reduced the continuous ranked probability score of WeatherNext 2 from 0.964 to 0.687 K, while raising its forecast-to-observed record frequency ratio from 0.370 to 1.970. Applied to Pangu-Weather forecasts, spatially conditioned tail-aware calibration (STAC) gave a ratio of 1.070. Its record Brier score was lower than that of WeatherNext 2 with EMOS by $2.520\times10^{-4}$ (95\% block-bootstrap interval for the difference, $[-3.140,-1.910]\times10^{-4}$). A reduction in loss on non-exceedances outweighed an increase in loss on exceedances. At the 99th percentile, the interval for the total score difference included zero. For the record alone, the block-bootstrap interval for the Brier score difference between a rolling isotonic correction and STAC included zero. At days 8--10, WeatherNext 2 with EMOS retained higher discrimination, with an area under the receiver operating characteristic curve of 0.965 compared with 0.889 for STAC. These results show why the overall temperature score, the probabilities assigned to rare events and the ability to rank those events merit separate assessment.
\end{abstract}


\section{Introduction}

A medium-range forecast of extreme heat can answer several different questions. A point forecast estimates an expected temperature or another feature of its distribution. A probability forecast estimates the chance of crossing a specified level at a location and lead time. For a threshold as high as a local historical record, a small redistribution of probability in the upper tail can substantially change the expected frequency of exceedances. An aggregate score for the temperature distribution can improve even when that frequency deteriorates. The distinction matters when a forecast product is judged using a full-distribution score and subsequently interpreted through its probabilities of rare events.

Previous studies have reached different conclusions about extreme-heat performance because they target different forecast quantities and verification settings. Zhang et al. (2026) reported that ECMWF's high-resolution forecast had lower errors than the tested deterministic machine-learning forecasts on realised record exceedances. Conditioning the verification sample on realised extremes changes the question answered by a score and can favour forecasts that issue extremes too frequently (Lerch et al., 2017). Wessel et al. (2026) assessed probabilistic heat forecasts and found advantages for native machine-learning ensembles over the IFS ensemble, including an EMOS comparator and a station-based check. Van Velthoven et al. (2026) further examined probabilities above progressively deeper thresholds and separated contributions from exceedances and non-exceedances. Biegert, Allen, Zhang and Lerch (2026) showed how evaluation choices can change system rankings for extremes. Together these studies motivate examining the forecast distribution, the exceedance probability and the verification protocol explicitly.

Post-processing adds a further question. Gaussian ensemble model output statistics estimates a distribution by calibrating forecast location and spread (Gneiting et al., 2005). Related operational statistical workflows have improved probabilistic forecasts derived from machine-learning models (Trotta et al., 2025). The improvement in a distribution-wide score, however, leaves open how the specific correction affects record probabilities. Tail calibration concerns properties that can be concealed by pooled scores or marginal event frequencies (Allen et al., 2025). A forecast with the correct aggregate frequency can still assign too much probability to some cases and too little to others. Forecast guidance also depends on discrimination, which measures how well forecasts rank cases that exceed the threshold above those that do not.

We investigate how post-processing changes probabilities of exceeding a fixed historical monthly maximum of daily temperature at individual land grid cells. We first compare raw and post-processed Pangu-Weather, IFS-ENS and WeatherNext 2 on a common archive. We then compare Gaussian EMOS and spatially conditioned tail-aware calibration (STAC) on the same Pangu-Weather backbone and use post-processed WeatherNext 2 as a strong cross-system comparator. We examine how frequency bias and the Brier score contributions from exceedances and non-exceedances change as the threshold rises, and whether a rolling correction for the historical record alone closes the score difference. Finally, we assess how discrimination changes with lead time and use the available ablations, a change of Pangu initialisation source and a retrospective 2020 comparison to delimit the findings.

\section{Data and methods}

\subsection{Forecast archives, verification domain and target variable}

We used the conservatively regridded 1.5\ensuremath{{}^\circ} products on the 121 \ensuremath{\times} 240 regular latitude--longitude grid of WeatherBench 2 (Rasp et al., 2024). The main verification reference was the ECMWF Reanalysis v5 (ERA5, Hersbach et al., 2020). The verification domain comprised 6,693 grid cells between 60\ensuremath{{}^\circ}S and 80\ensuremath{{}^\circ}N with a land fraction greater than 0.5. We weighted these cells by the cosine of latitude for all principal scores except the area under the receiver operating characteristic curve. The main comparison used 292 daily initialisations at 00 UTC from 15 March to 31 December 2022. The common start date allowed the rolling post-processors to accumulate history, including for WeatherNext 2, whose archive in this study begins in January 2022.

At each grid cell we defined daily maximum temperature, $T_{\mathrm{x}}$, as the maximum of the 2 m temperature at 06, 12 and 18 UTC on the valid date and 00 UTC on the following date. For an initialisation on date $t$ at 00 UTC, forecast day $d$ verified on date $t+d-1$, and the last sample in the window was at lead $24d$ hours. We formed the same four-sample maximum for the reference and for each forecast member before calculating scores. Thus $T_{\mathrm{x}}$ represents a UTC-based, four-sample daily maximum on the verification grid. The primary lead group was days 1--7, with days 8--10 used to assess longer leads. The initialisation window was held fixed across lead times, so the late-December forecasts could verify in early January.

The raw forecasts were deterministic Pangu-Weather (Bi et al., 2023), the 50-member ECMWF Integrated Forecasting System ensemble (IFS-ENS), and the 64-member WeatherNext 2 ensemble (FGN; Alet et al., 2025). In tables and figures, FGN denotes WeatherNext 2. The forecast archives are used as distributed through WeatherBench 2; their training periods are documented in the model papers. The principal Pangu archive was initialised from ERA5. We also evaluated a separate Pangu archive initialised from ECMWF operational high-resolution analysis fields (HRES). The IFS-ENS and FGN results describe the specific WeatherBench 2 archives used here. The inter-system comparison combines differences in both forecast systems and their initialisation products.

\subsection{Fixed historical thresholds}

Let $R_{17}(g,m)$ be the largest ERA5 $T_{\mathrm{x}}$ at grid cell $g$ in calendar month $m$ over 1979--2017. For each forecast case we selected $m$ from the valid date. An event occurred when the verifying $T_{\mathrm{x}}$ strictly exceeded $R_{17}(g,m)$. The value of $R_{17}$ remained fixed throughout evaluation. Each exceedance therefore represents a day above the historical monthly maximum at that grid cell. Different initialisations and leads can verify against the same grid-cell day.

To compare events across the upper tail, we also evaluated the monthly 99th percentile $q_{99}(g,m)$, $R_{17}+0.5\sigma_m$ and $R_{17}+1.0\sigma_m$. The quantity $\sigma_m(g)$ is the standard deviation over all baseline days in calendar month $m$ after subtracting the smoothed day-of-year climatological mean at that grid cell. We used the same definition of $\sigma_m$ and the same pooling over leads for the entire threshold comparison. The shared baseline period follows the historical record construction in Zhang et al. (2026). The target and lead verification here use the four-sample daily maximum. The multithreshold comparison also follows the recent use of explicit exceedance depths in probabilistic verification by Van Velthoven et al. (2026).

\subsection{Gaussian ensemble model output statistics}

We evaluated a static Pangu post-processor and a rolling formulation for Pangu, IFS-ENS and FGN. Both formulations used Gaussian ensemble model output statistics (EMOS) and minimised the closed-form continuous ranked probability score (CRPS) of the fitted Gaussian distribution (Gneiting et al., 2005). The inclusion of both raw and post-processed IFS forecasts follows the design of Wessel et al. (2026). Here we also apply the rolling Gaussian correction to the native FGN ensemble.

The static Pangu EMOS fitted coefficients separately at each grid cell and lead in standardised anomaly space. If $x$ denotes the current deterministic forecast and $s_{\rm lag}$ the standard deviation across the available current and 12- and 24-hour lagged forecasts for the same valid date, the mean was $a+bx$ and the standard deviation was $\exp\{c+e\log(s_{\rm lag}+0.05)\}$. We used initialisations from 2018--2020 for the main 2022 fit. Forecasts were returned to physical units for verification.

The rolling formulation fitted a Gaussian with mean $a+b\bar{x}$ and variance $c^2+e^2s^2$ in physical units. For FGN and IFS-ENS, $\bar{x}$ and $s^2$ were the raw ensemble mean and sample variance. For Pangu, the location predictor was the current deterministic forecast and the spread predictor was the sample variance of the available same-day lagged forecasts, including the current forecast and lags at 12-hour intervals up to 72 hours. Each fit pooled the target cell and available land neighbours in a 3 \ensuremath{\times} 3 window. For an initialisation $t$ at lead day $d$, the training initialisations ranged from $t-d-60$ to $t-d-1$, inclusive. The latest verifying window in that fit had ended before $t$. These are retrospective fits with a cutoff based on the valid time of archived ERA5. The evaluation did not measure real-time dissemination delays. Each Gaussian EMOS distribution was represented in the verification by its 50 equally spaced mid-quantiles, $(j-0.5)/50$ for $j=1,\ldots,50$.

\subsection{Spatially conditioned tail-aware calibration}

Spatially conditioned tail-aware calibration (STAC) is a learned spatial post-processor that generates ensemble members from Pangu fields. The predictive model is trained jointly on daily maximum temperature, daily minimum temperature and daily maximum 10 m wind speed. Its predictors describe the current and lagged forecasts for the same valid date, initialisation-time land and surface state, static geographical and climatological fields, and monthly climatological context. The model also receives lead, season and initialisation-hour information. It generates spatial draws from a Gaussian latent vector and combines a learned rescaling of the current standardised Pangu forecast with additive corrections. Forecasts at different lead days are generated separately. The architecture uses spatial conditioning in the style of neural distributional post-processing (Pic et al., 2025).

For the weather targets we standardised by grid-cell and day-of-year ERA5 means and standard deviations derived from 1979--2017. The training objective combines an almost-fair CRPS for the full predictive distribution, an almost-fair CRPS applied to tail-transformed forecasts and verifying values, and a patch energy score (Lang et al., 2026). The tail transform emphasises the upper tail for $T_{\mathrm{x}}$ and wind speed and the lower tail for minimum temperature. The two CRPS terms give land points a larger weight than ocean points at the same latitude. The full channel specification, loss weights and optimisation settings are not publicly available.

The main models were trained on initialisations from 2018--2020, and the checkpoint was selected on 00 UTC initialisations from 2021. Three training seeds produced separate networks, and each forecast case uses a 50-member ensemble formed from their draws. Forecasts were verified after conversion to physical units. For the HRES-initialised Pangu experiment we reused the fitted networks and the ERA5 ancillary inputs with forecasts from the operationally initialised Pangu archive. This tests sensitivity to the archive used to initialise the backbone and remains a hindcast experiment. The trained weights are not publicly available.

\subsection{Single-threshold recalibration and sensitivity experiments}

We tested whether a separate correction limited to $R_{17}$ could recover the record score of FGN with rolling EMOS. We fitted an area-weighted isotonic map of forecast probability to observed frequency separately for each lead and initialisation. We pooled land grid cells over the corresponding 60-initialisation training window ending at $t-d-1$, including eligible pre-evaluation forecasts, and clipped predictions outside the range of fitted probabilities to the nearest fitted endpoint. We then applied the map to each current grid-cell probability and computed its Brier loss before spatial averaging. This procedure issues only an $R_{17}$ probability correction. It used the same 2022 initialisations and days 1--7 as the primary record comparison.

We assessed the effects of two model components using three-seed ensembles of the same size as the full model. One omitted the tail-weighted term from the training loss. The other set the land and surface fields and the preceding-window ERA5 fields to zero. This second intervention removed a group of predictors. Exploratory single-seed checks also removed the monthly context channels or used a Gaussian output head.

For the 2020 retrospective comparison, we trained STAC on initialisations in 2018, 2019 and 2021 and used the final checkpoint. Static Pangu EMOS was fitted on the same three initialisation years, and rolling IFS-ENS EMOS was retained as a separate benchmark. We verified 292 initialisations from 15 March to 31 December 2020, using days 1--7. The split is by initialisation year, so a small number of late-December 2020 forecasts verify in early January 2021. The experiment is a retrospective consistency check rather than a fully independent evaluation.

\subsection{Verification and uncertainty}

For a threshold $u$ at each case $i$ (grid cell, initialisation and lead), the forecast probability $p_i$ was the fraction of members with $T_{\mathrm{x}}>u$, and the event indicator was $o_i=\mathbf{1}(T_{\mathrm{x},i}^{\mathrm{ERA5}}>u)$. Raw Pangu was evaluated as a point mass. We verified both exceedances and non-exceedances for all event scores, consistent with evaluating extreme-event probabilities over the full sample (Lerch et al., 2017). Using area weights $w_i$, the forecast-to-observed occurrence ratio was $Q=\sum_i w_i p_i/\sum_i w_i o_i$. A value of one denotes matching aggregate probability mass and observed event frequency. Conditional reliability was assessed separately by comparing weighted mean forecast probabilities with observed event frequencies in probability bins.

We evaluated the full $T_{\mathrm{x}}$ distribution with the ensemble CRPS using the fair correction of Ferro (2014). For a point forecast this reduces to absolute error. We applied the same finite-ensemble expression to native ensembles, generated members and EMOS mid-quantiles. The fair-score sampling interpretation applies to random ensemble draws, and the EMOS mid-quantiles are deterministic. For the record-threshold-weighted CRPS (twCRPS), we used the same score after applying $v(x)=\max(x,R_{17})$ to forecast values and the verifying value, so differences below the threshold make no direct contribution (Allen et al., 2023). Both scores are in kelvin.

The ordinary Brier score is the weighted mean squared probability error (Brier, 1950). We partitioned it by verification outcome, following the outcome-based diagnostics of Van Velthoven et al. (2026).

\[
\mathrm{BS}=\frac{\sum_{i:o_i=1}w_i(1-p_i)^2}{\sum_i w_i}
+\frac{\sum_{i:o_i=0}w_i p_i^2}{\sum_i w_i}.
\]

The first term measures loss on exceedances and the second measures loss on non-exceedances. Both use the total sample weight in the denominator. The calculation uses the issued probability for each case, without applying a warning probability threshold.

For discrimination we computed the area under the receiver operating characteristic curve (AUC) separately at each lead by pooling initialisations and grid cells with equal case weights. We broke equal-probability rankings by adding $10^{-6}\tanh((\bar{x}_i-R_{17,i})/(1\,\mathrm{K}))$ to each probability, where $\bar{x}_i$ is the ensemble mean. We averaged the lead-wise AUCs within each lead group. The CORP reliability decomposition (Dimitriadis et al., 2021) used area-weighted isotonic regression fitted and scored on the same verification cases for a given system and lead. We report the difference between the original Brier score and the score after this in-sample recalibration, $\mathrm{MCB}=\mathrm{BS}-\mathrm{BS}_{\rm rc}$. At a common verifying event frequency, lower $\mathrm{BS}_{\rm rc}$ indicates more resolution in this diagnostic. This fit has a different purpose from the historical-window recalibration in Section 2.5.

In the main comparison we averaged the lead-wise CRPS, twCRPS, BS and AUC within each lead group. In the multithreshold analysis we pooled weighted contributions over grid cells, initialisations and leads before forming $Q$, the Brier terms and the probability-bin diagnostics. We obtained pointwise 95\% percentile intervals from 1,000 bootstrap replicates using seven-day blocks of initialisation dates, carrying along each date's spatial field and leads. Paired differences use the same resampled dates for both systems. For multithreshold scores we used non-overlapping blocks defined from the beginning of the evaluation period, including a shorter final block. For the designated record-only comparisons, ablations, online comparison and long-lead AUC contrast, we used circular moving blocks. We retained the original grid-cell rank comparisons when resampling AUC. Intervals describe variability across initialisations for the fitted forecast systems.

For the probability-bin diagnostics, we used fixed bin edges on the unit interval, with finer bins at low probabilities. To examine spatial variation, we classified grid cells into quintiles of the annual mean monthly gap $R_{17}-q_{99}$ divided by the annual mean day-of-year climatological standard deviation. The smallest gap defines the shortest-tail group. At lead day 7 this diagnostic used every other initialisation and gave grid cells equal weight within each group. Intervals for selected ratios retained that sampling pattern on the calendar-day bootstrap grid. In a separate diagnostic, we grouped cells by the within-initialisation 0.33 and 0.67 quantiles of ERA5 upper-layer soil moisture at 00 UTC. We area-weighted within each group and averaged forecast and observed frequency across the sampled initialisations before taking their ratio. These partitions describe associations in the archive.

\section{Results}

\subsection{Changes in overall temperature skill and record probabilities}

All three Gaussian rolling post-processors reduced the days 1--7 full-distribution CRPS relative to their raw forecasts, while each increased the total probability assigned to $R_{17}$ exceedances (Table 1 and Figure 1). WeatherNext 2 (FGN) CRPS decreased from 0.964 to 0.687 K, the larger reduction of the two native ensembles. Its occurrence ratio increased from 0.370 (95\% CI 0.321--0.416) to 1.967 (1.788--2.189), passing from underprediction to overprediction. IFS-ENS was close to the observed frequency before EMOS ($Q=1.091$) and forecast 2.809 times the observed frequency after rolling EMOS. For Pangu, $Q$ rose from 0.673 to 1.852 under static EMOS and to 2.927 under rolling EMOS. In particular, the IFS-ENS rolling correction improved overall CRPS while the $R_{17}$ Brier score increased.

With the same Pangu archive as the static EMOS baseline, STAC minus static EMOS gave a days 1--7 $R_{17}$ Brier score difference of $-3.060\times10^{-4}$ (95\% CI $[-3.650,-2.550]\times10^{-4}$). It also had a lower point estimate of record twCRPS. WeatherNext 2 with rolling EMOS achieved a lower full-distribution CRPS and higher record AUC than STAC. STAC achieved the lower record Brier score of the two. Its occurrence ratio was 1.074 (1.009--1.162), indicating residual aggregate overprediction at $R_{17}$.

\begin{table}[!htbp]
\centering
\caption{Days 1--7 $T_{\mathrm{x}}$ verification over land for 15 March--31 December 2022 initialisations. The threshold is $R_{17}$ for twCRPS, BS, AUC and $Q$. CRPS, twCRPS, BS and AUC are means of lead-wise values. In the $Q$ column, parentheses give 95\% intervals; fixed initialisation blocks are used for the systems in Table 2 and for the raw systems, and moving blocks for rolling Pangu EMOS. AUC uses equal case weights. The other verification measures use area weights. FGN denotes WeatherNext 2. Pooled values in the text and in Tables 3 and 5 can differ in the final digit because this table reports lead-wise means.}
\small
\begin{tabularx}{\textwidth}{@{}>{\raggedright\arraybackslash}Xrrrrr@{}}
\toprule
Forecast system & CRPS (K) & twCRPS ($10^{-3}$ K) & BS ($10^{-4}$) & AUC & $Q$ (95\% CI) \\
\midrule
Pangu, raw & 1.163 & 2.290 & 29.730 & 0.978 & 0.673 (0.549--0.790) \\
Pangu + static EMOS & 0.793 & 1.720 & 24.270 & 0.980 & 1.852 (1.663--2.106) \\
Pangu + rolling EMOS & 0.796 & 2.340 & 29.660 & 0.984 & 2.927 (2.608--3.332) \\
IFS-ENS, raw & 1.023 & 2.000 & 28.190 & 0.980 & 1.091 (0.921--1.319) \\
IFS-ENS + rolling EMOS & 0.815 & 2.140 & 30.010 & 0.989 & 2.809 (2.402--3.290) \\
FGN, raw & 0.964 & 1.770 & 25.010 & 0.986 & 0.370 (0.321--0.416) \\
FGN + rolling EMOS & 0.687 & 1.600 & 23.760 & 0.992 & 1.967 (1.788--2.189) \\
STAC & 0.746 & 1.520 & 21.210 & 0.983 & 1.074 (1.009--1.162) \\
STAC with HRES-initialised Pangu & 0.774 & 1.600 & 22.140 & 0.982 & 1.116 (1.024--1.239) \\
\bottomrule
\end{tabularx}
\end{table}

\begin{figure}[htbp]
\centering
\includegraphics[height=225pt]{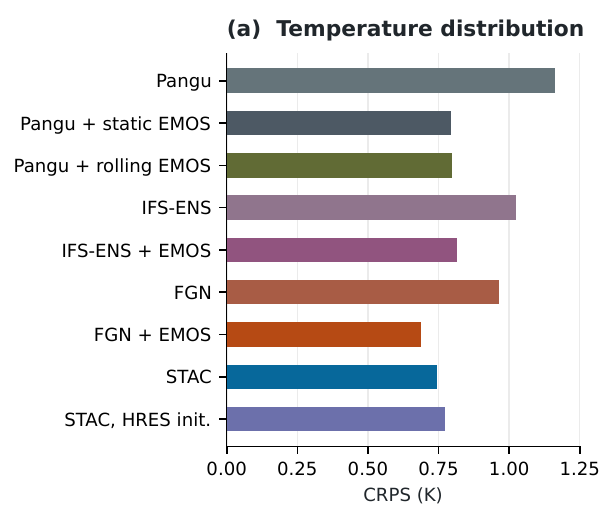}\hfill
\includegraphics[height=225pt]{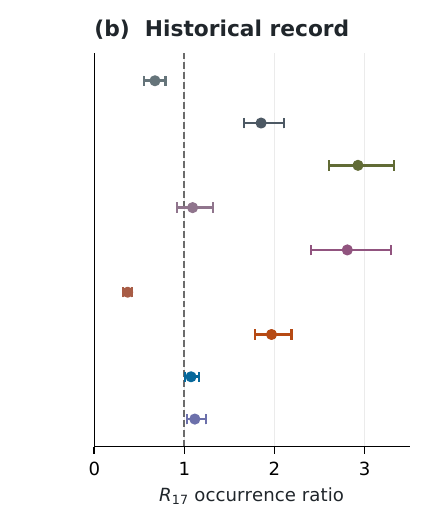}\\[4pt]
\includegraphics[width=0.55\textwidth]{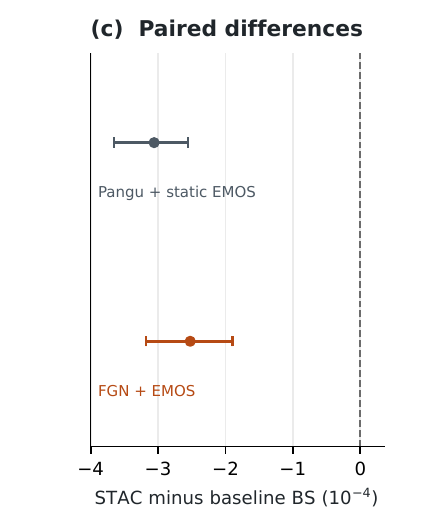}
\caption{Changes in overall temperature performance and record probability after post-processing. (a) Mean days 1--7 $T_{\mathrm{x}}$ CRPS in kelvin for raw Pangu, IFS-ENS and FGN and their post-processed variants. (b) The corresponding area-weighted $R_{17}$ occurrence ratios, with a reference line at $Q=1$. Error bars give the system-specific 95\% intervals and use the block schemes in Table 1. (c) Paired record Brier score differences for STAC minus static Pangu EMOS and STAC minus rolling FGN EMOS, with 95\% circular-block intervals. Panels use 2022 land forecasts from 15 March--31 December.}
\end{figure}

\subsection{Threshold depth and the sources of Brier score differences}

The frequency discrepancies increased with threshold depth for all three EMOS baselines in Table 2. For WeatherNext 2 + EMOS, $Q$ rose from 1.325 at $q_{99}$ to 6.714 at $R_{17}+1.0\sigma_m$. Static Pangu EMOS reached 3.467 at the deepest threshold and IFS-ENS + rolling EMOS reached 12.480. STAC's occurrence ratios remained near one across the four pooled point estimates, and the uncertainty grew at the deeper thresholds. At $R_{17}+1.0\sigma_m$, there were 654 verifying exceedance cases and the interval on $Q$ for STAC extended from 0.590 to 1.768. The cases are counted at grid-cell, initialisation and lead level and can share a verifying day. At $R_{17}$, forecasts in the 6--10\% probability bin had a weighted mean probability of 8.0\% for both systems. The verifying frequency in that bin was 6.0\% for STAC and 2.4\% for WeatherNext 2 + EMOS (Figure 2). The two systems assigned different cases to this bin.

\begin{figure}[htbp]
\centering
\includegraphics[height=18.5pt]{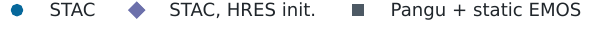}\hfill
\includegraphics[height=18.5pt]{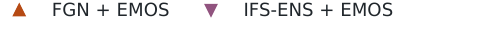}\\[2pt]
\includegraphics[height=224pt]{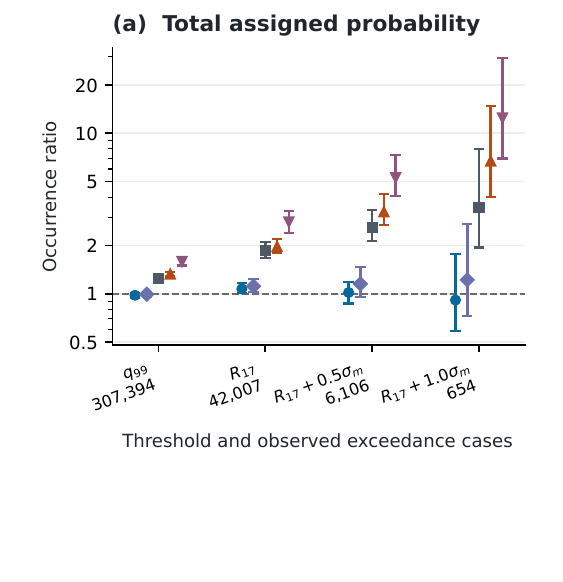}\hfill
\includegraphics[height=224pt]{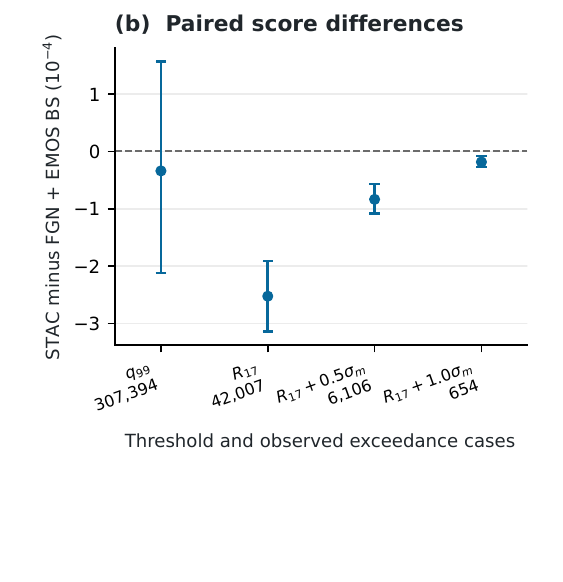}\\[4pt]
\includegraphics[width=0.60\textwidth]{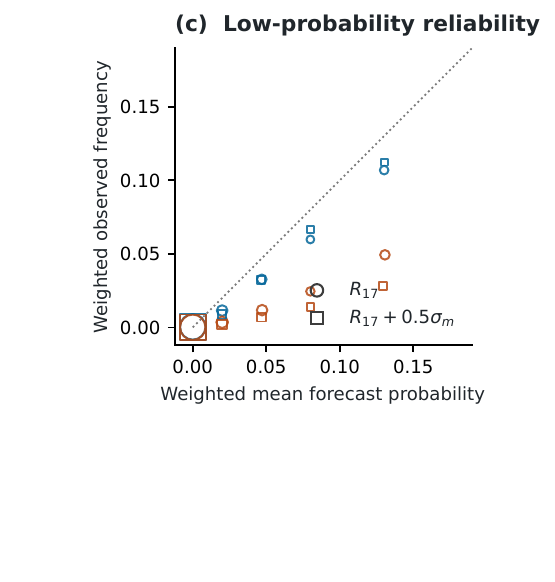}
\caption{Probability bias and reliability as the exceedance threshold increases. (a) Days 1--7 pooled occurrence ratios for STAC, STAC with HRES-initialised Pangu and the Pangu, FGN and IFS-ENS EMOS baselines. The thresholds are $q_{99}$, $R_{17}$, $R_{17}+0.5\sigma_m$ and $R_{17}+1.0\sigma_m$. Numbers along the horizontal axis give the observed exceedance case counts in Table 2. (b) Pooled BS difference between STAC and FGN + EMOS at the same thresholds. Bars in (a) and (b) are 95\% fixed-block intervals. (c) Area-weighted reliability of STAC and FGN + EMOS at $R_{17}$ and $R_{17}+0.5\sigma_m$. Each point gives the weighted forecast probability and observed frequency within the same bin. Marker areas increase with the square root of each bin's share of the total case weight, and the diagonal denotes perfect agreement. The panel focuses on low probabilities. All panels use the same monthly anomaly standard deviation and pooled 2022 cases.}
\end{figure}

\begin{table}[!htbp]
\centering
\caption{Occurrence ratios by threshold, pooled over the 2022 land sample at days 1--7. Each threshold is evaluated at 13,680,492 grid-cell--initialisation--lead cases. The observed count is unweighted. The ratios use area weights. Parentheses give 95\% fixed-block intervals. FGN denotes WeatherNext 2; FGN and IFS-ENS use rolling EMOS, and Pangu uses static EMOS.}
\small
\resizebox{\textwidth}{!}{\begin{tabular}{@{}lrcccc@{}}
\toprule
Threshold & Events & STAC & FGN + EMOS & Pangu + EMOS & IFS-ENS + EMOS \\
\midrule
$q_{99}$ & 307,394 & 0.979 (0.949--1.013) & 1.325 (1.282--1.369) & 1.252 (1.194--1.318) & 1.595 (1.502--1.692) \\
$R_{17}$ & 42,007 & 1.074 (1.009--1.162) & 1.967 (1.788--2.189) & 1.852 (1.663--2.106) & 2.809 (2.402--3.290) \\
$R_{17}+0.5\sigma_m$ & 6,106 & 1.020 (0.871--1.189) & 3.239 (2.671--4.184) & 2.583 (2.122--3.313) & 5.303 (4.069--7.294) \\
$R_{17}+1.0\sigma_m$ & 654 & 0.914 (0.590--1.768) & 6.714 (4.035--14.862) & 3.467 (1.943--7.972) & 12.480 (6.968--29.584) \\
\bottomrule
\end{tabular}}
\end{table}

At $R_{17}$, the pooled BS was $21.220\times10^{-4}$ for STAC and $23.750\times10^{-4}$ for WeatherNext 2 (FGN) + EMOS. The improvement was attributable to the lower loss on non-exceedances ($5.090$ versus $11.510\times10^{-4}$). The loss contribution on exceedances was higher for STAC ($16.140$ versus $12.240\times10^{-4}$). The paired differences were $-6.420\times10^{-4}$ on non-exceedances and $+3.900\times10^{-4}$ on exceedances, giving a net difference of $-2.520\times10^{-4}$ (Table 3 and Figure 3). Against the same-backbone static Pangu EMOS, the estimated differences in the two components were $-2.930$ (95\% CI $[-3.590,-2.360]$) on non-exceedances and $-0.130$ ($[-0.720,0.320]$) on exceedances, in $10^{-4}$ units.

The component trade-off against WeatherNext 2 + EMOS was also present at $q_{99}$, where the interval for the total BS difference included zero. At both thresholds above $R_{17}$, the paired intervals for the net differences were below zero, with smaller absolute differences at each successive threshold. All of these comparisons are based on the 2022 sample.

\begin{table}[!htbp]
\centering
\caption{STAC minus WeatherNext 2 (FGN) + rolling EMOS differences in the Brier score and its two outcome contributions for the 2022 land sample at days 1--7. Values and 95\% paired fixed-block intervals are in $10^{-4}$ units. Negative differences indicate lower loss for STAC. FGN denotes WeatherNext 2.}
\small
\begin{tabularx}{\textwidth}{@{}>{\raggedright\arraybackslash}Xrrr@{}}
\toprule
Threshold & Exceedance contribution & Non-exceedance contribution & Total BS \\
\midrule
$q_{99}$ & $+21.100$ [17.400, 24.950] & $-21.440$ [-24.840, -17.970] & $-0.340$ [-2.110, 1.570] \\
$R_{17}$ & $+3.900$ [3.050, 4.760] & $-6.420$ [-7.380, -5.510] & $-2.520$ [-3.140, -1.910] \\
$R_{17}+0.5\sigma_m$ & $+0.720$ [0.470, 1.030] & $-1.560$ [-1.830, -1.320] & $-0.840$ [-1.080, -0.560] \\
$R_{17}+1.0\sigma_m$ & $+0.110$ [0.040, 0.220] & $-0.290$ [-0.360, -0.230] & $-0.180$ [-0.270, -0.080] \\
\bottomrule
\end{tabularx}
\end{table}

\begin{figure}[htbp]
\centering
\includegraphics[height=218pt]{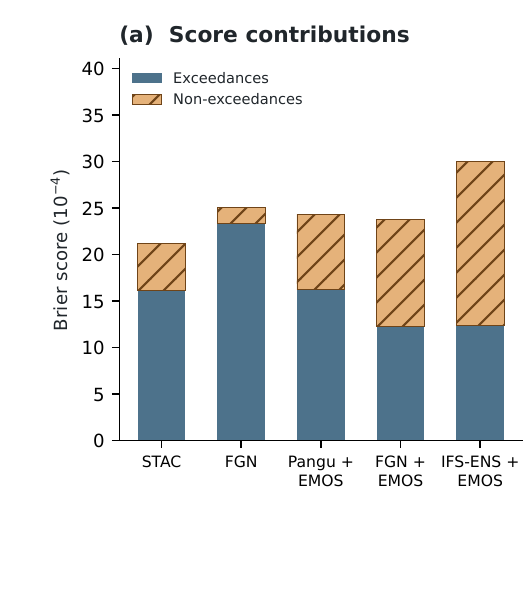}\hfill
\includegraphics[height=218pt]{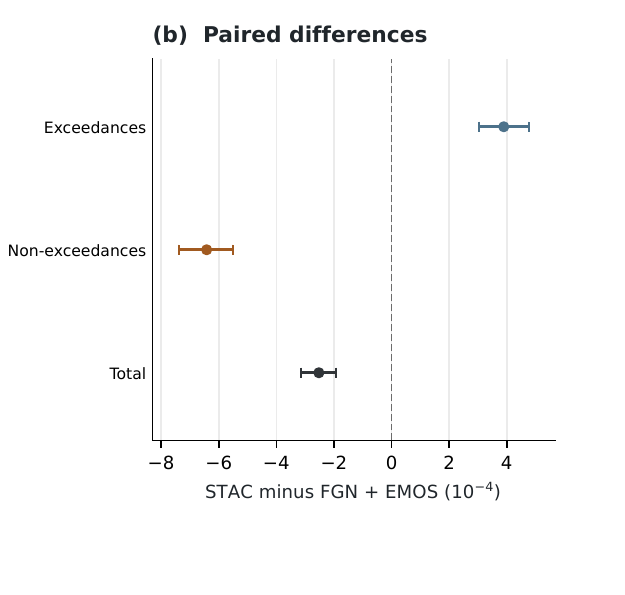}\\[4pt]
\includegraphics[width=0.55\textwidth]{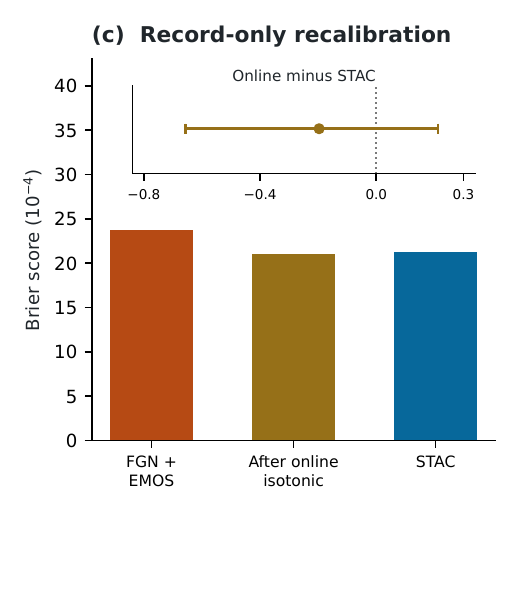}
\caption{Outcome contributions to the record Brier score and the result of online recalibration. (a) Area-weighted exceedance and non-exceedance contributions to $R_{17}$ BS for STAC, raw FGN and the three EMOS comparators. (b) STAC minus FGN + EMOS differences in the two contributions and their sum, with 95\% paired fixed-block intervals. (c) $R_{17}$ BS for FGN + EMOS before and after rolling isotonic recalibration and for STAC on the same cases. The inset shows the online-minus-STAC difference with its 95\% paired circular-block interval. Panels use days 1--7 and the 2022 land sample with 42,007 exceedance cases. Scores are in units of $10^{-4}$.}
\end{figure}

\subsection{Online recalibration of the record probability}

The rolling isotonic map reduced the WeatherNext 2 (FGN) + EMOS $R_{17}$ score from $23.750\times10^{-4}$ to $21.030\times10^{-4}$ and brought $Q$ from 1.967 to 0.972. The STAC score on the same cases was $21.220\times10^{-4}$, with $Q=1.074$. The paired difference (online recalibration minus STAC) was $-0.200\times10^{-4}$, with a 95\% interval of $[-0.660,0.210]\times10^{-4}$ (Figure 3). The interval includes zero. The comparison addresses the single event for which we trained the isotonic map.

\subsection{Discrimination at longer lead times}

At days 8--10 (Table 4), STAC retained an occurrence ratio of 1.090 while WeatherNext 2 (FGN) + rolling EMOS had a ratio of 2.730. The corresponding record AUCs were 0.889 and 0.965, a paired difference of $-0.076$ (95\% CI $[-0.097,-0.056]$; Figure 4). The raw WeatherNext 2 and IFS-ENS ensembles also ranked exceedances above non-exceedances more effectively than STAC on this measure, with AUCs of 0.945 and 0.930, respectively. Full-distribution CRPS over these leads was 1.417 K for STAC and 1.242 K for WeatherNext 2 + EMOS.

In the CORP analysis, STAC had a smaller miscalibration contribution at both day 7 and day 10 (1.0\% and 1.8\% of BS, compared with 11.7\% and 13.6\% for WeatherNext 2 + EMOS). WeatherNext 2 + EMOS achieved lower $\mathrm{BS}_{\rm rc}$ at those leads. The values for WeatherNext 2 + EMOS and STAC were $23.940$ and $25.450\times10^{-4}$ at day 7, and $28.040$ and $29.470\times10^{-4}$ at day 10 (Figure 4). Thus WeatherNext 2 + EMOS retained better separation between events and non-events under both diagnostics at the longer leads. The comparison spans different backbones, predictor sets and post-processors, so it locates the discrimination gap between the tested systems.

\begin{figure}[htbp]
\centering
\includegraphics[height=18.5pt]{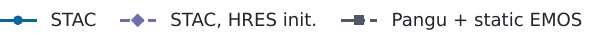}\hfill
\includegraphics[height=18.5pt]{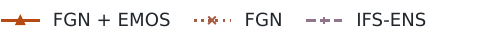}\\[2pt]
\includegraphics[height=190pt]{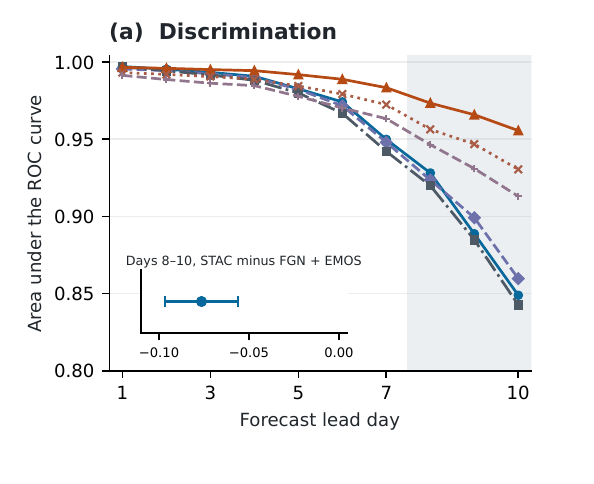}\hfill
\includegraphics[height=190pt]{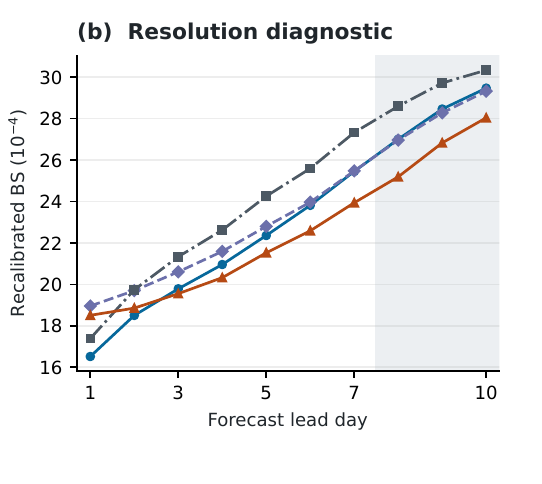}\\[4pt]
\includegraphics[width=0.60\textwidth]{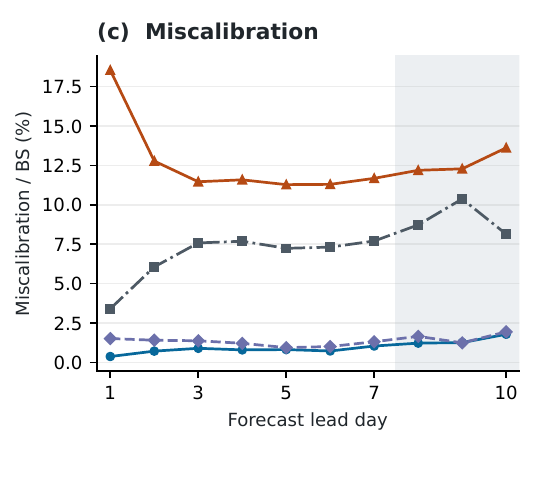}
\caption{Lead dependence of discrimination and reliability for $R_{17}$. (a) Equal-case-weight AUC by lead day for STAC, the HRES-initialised STAC variant, static Pangu EMOS, FGN + EMOS and the raw FGN and IFS-ENS ensembles. (b) $\mathrm{BS}_{\rm rc}$ in units of $10^{-4}$ and (c) MCB as a percentage of BS at each lead for STAC and the post-processed comparators. Panels (b) and (c) come from area-weighted, in-sample isotonic fits. Daily curves are point estimates. The inset in (a) shows the mean day-8--10 AUC difference for STAC minus FGN + EMOS and its 95\% paired circular-block interval. The sample comprises the 2022 land initialisations.}
\end{figure}

\begin{table}[!htbp]
\centering
\caption{Days 8--10 record verification for the 2022 land sample. CRPS is the mean lead-wise value. $Q$ and AUC are formed as in Table 1. Point estimates are shown; the paired AUC difference and its interval are given in the text. FGN denotes WeatherNext 2.}
\small
\begin{tabular}{@{}lrrr@{}}
\toprule
Forecast system & CRPS (K) & $Q$ & AUC \\
\midrule
STAC & 1.417 & 1.090 & 0.889 \\
FGN + rolling EMOS & 1.242 & 2.730 & 0.965 \\
\bottomrule
\end{tabular}
\end{table}

\subsection{Ablations and spatial variation}

The tail-weighted training term reduced overprediction most clearly at longer leads. Full STAC had $Q=1.090$ at days 8--10, compared with 1.340 when trained without that term. At days 1--7 the ratios were 1.070 and 1.140. The paired BS intervals for this ablation included zero in both lead groups (Table 5). Removing the land, surface and preceding-window inputs increased BS at days 1--7, and its effect on $Q$ was unresolved. At days 8--10 both paired intervals for this input-group comparison included zero. The exploratory single-seed networks with the Gaussian head and with the monthly context channels suppressed produced $Q$ values of 1.060 and 0.990, respectively, compared with 1.010 for the full single-seed configuration. These checks leave the individual contributions of the spatial architecture, output distribution and monthly context unresolved.

The domain-average occurrence ratios masked geographical variation. At day 7 in the shortest-tail quintile, STAC still overpredicted $R_{17}$ exceedances ($Q=1.710$, 95\% CI 1.320--2.260). The corresponding values were 3.410 for static Pangu EMOS and 2.320 (1.910--2.930) for WeatherNext 2 (FGN) + EMOS (Figure 5). At day 7, static Pangu EMOS had $Q=1.990$ and 3.000 in the driest and wettest soil-moisture groups, while STAC yielded 1.130 and 1.270. The physical interpretation remains open because the strata can differ in regional climate, seasonal conditions and the local definition of extremes. The input ablation changes several predictors jointly and addresses their combined predictive effect.

\begin{table}[!htbp]
\centering
\caption{Area-weighted $R_{17}$ verification for full STAC and its paired three-seed ablations in 2022. Each difference is full STAC minus the indicated ablation. Brackets give 95\% paired circular-block intervals.}
\small
\begin{tabularx}{\textwidth}{@{}l>{\raggedright\arraybackslash}Xrrrr@{}}
\toprule
Lead days & Ablation & Full $Q$ & Ablated $Q$ & $\Delta Q$ & $\Delta$BS ($10^{-4}$) \\
\midrule
1--7 & Tail term omitted & 1.070 & 1.140 & -0.070 [-0.090, -0.050] & -0.120 [-0.290, 0.050] \\
8--10 & Tail term omitted & 1.090 & 1.340 & -0.260 [-0.310, -0.210] & -0.190 [-0.440, 0.110] \\
1--7 & Land/surface inputs omitted & 1.070 & 1.100 & -0.020 [-0.060, 0.010] & -0.690 [-0.890, -0.480] \\
8--10 & Land/surface inputs omitted & 1.090 & 1.090 & -0.010 [-0.060, 0.050] & +0.090 [-0.260, 0.510] \\
\bottomrule
\end{tabularx}
\end{table}

\subsection{Initialisation-source transfer and retrospective 2020 comparison}

In the 2022 archive with HRES-initialised Pangu, the same trained STAC networks yielded $Q=1.116$ (1.024--1.239). Relative to WeatherNext 2 (FGN) + EMOS, the record BS difference was $-1.600\times10^{-4}$ (95\% paired CI $[-2.320,-0.800]\times10^{-4}$). This contrast retains ERA5 as the verification reference (Table 1 and Figure 5).

In the separate 2020 experiment, the retrospectively trained STAC yielded $Q=1.030$, compared with 1.830 for static Pangu EMOS fitted on the same years, 0.700 for raw Pangu and 4.890 for rolling IFS-ENS EMOS. The days 1--7 mean record BS was $27.410\times10^{-4}$ for STAC and $31.570\times10^{-4}$ for the matched static baseline. The paired STAC-minus-EMOS difference was $-4.160\times10^{-4}$ (95\% CI $[-5.020,-3.430]\times10^{-4}$). Because some preprocessing statistics span the evaluation year and a few late-December targets fall in early 2021, this experiment is a consistency check rather than a fully independent evaluation. The WeatherNext 2 (FGN) archive used for the main results contains 2022 initialisations only.

\begin{figure}[htbp]
\centering
\includegraphics[width=0.58\textwidth]{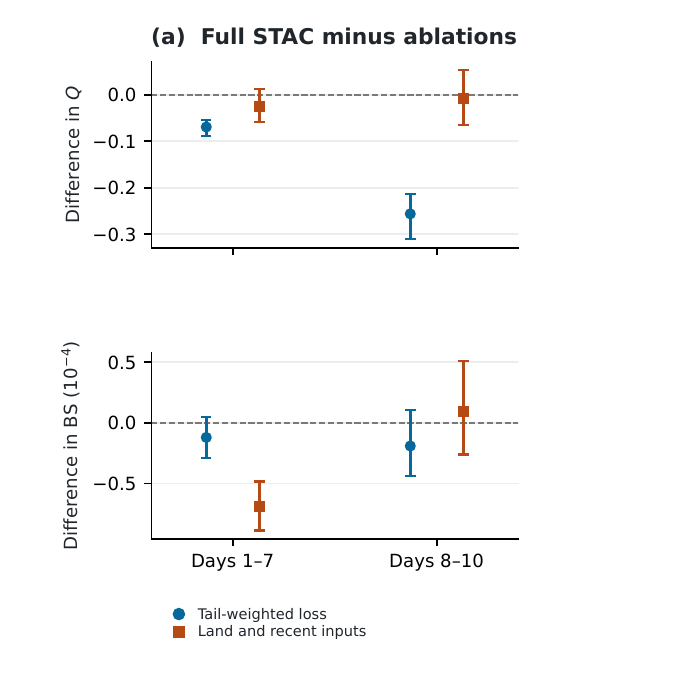}\\[4pt]
\begin{minipage}{0.9\textwidth}
\includegraphics[height=270pt]{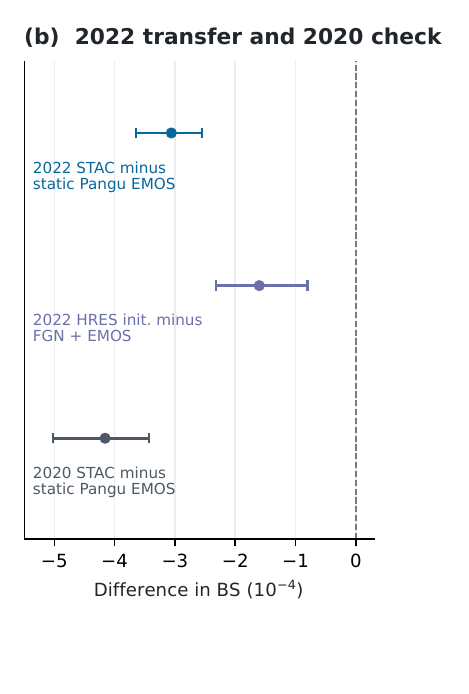}\hfill
\includegraphics[height=270pt]{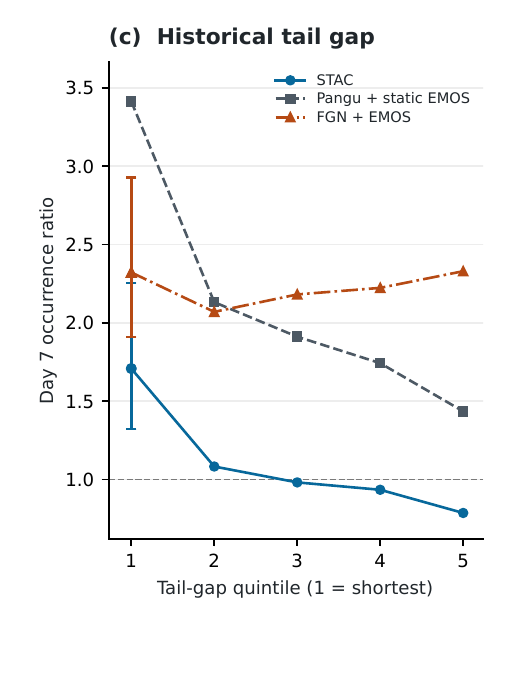}
\end{minipage}
\caption{Ablations, sensitivity to initialisation source and the shortest-tail locations. (a) Differences in $Q$ and $R_{17}$ BS for full STAC minus each three-seed ablation at days 1--7 and 8--10, with 95\% paired circular-block intervals. (b) Day-1--7 record BS differences for the 2022 principal ensemble against static Pangu EMOS, for the 2022 HRES-initialised variant against FGN + EMOS, and for the 2020 consistency ensemble against static Pangu EMOS fitted on the matched initialisation years. Each point shows the corresponding contrast, with a 95\% interval obtained by resampling its evaluation initialisations. (c) Day-7 $R_{17}$ occurrence ratio by quintile of the standardised historical tail gap for STAC, static Pangu EMOS and FGN + EMOS. This panel uses every other 2022 initialisation and equal grid-cell weights within groups. The shortest-tail STAC and FGN + EMOS points carry 95\% circular-block intervals on the calendar-day grid. Horizontal lines indicate zero for differences and one for occurrence ratios.}
\end{figure}

\section{Discussion}

\subsection{Why overall and record scores respond differently}

The change in WeatherNext 2 from underprediction to overprediction after Gaussian EMOS is a within-system result. The calibration objective rewards improvement across the entire predictive distribution. At a fixed high threshold, the forecast probability depends on how the fitted Gaussian combines its location and spread at that threshold. A useful correction to common temperatures can therefore assign too much mass beyond a record. The experiment identifies the net effect of the post-processor in these archives. The relative roles of the location adjustment, the estimated variance and the Gaussian tail shape require separate tests. The IFS-ENS comparison gives an especially clear example of the practical distinction, because the correction lowered CRPS while increasing record Brier loss.

The Brier decomposition identifies the cases responsible for the lower record loss of STAC. Relative to WeatherNext 2 with EMOS, the lower loss on non-exceedances outweighed the higher loss on exceedances. Against static Pangu EMOS, the reduction was again concentrated on non-exceedances, while the interval for the change on exceedances included zero. These results give different answers to whether the post-processor distributes probability better over the complete sample and whether it concentrates more probability on realised extremes. An operational comparison of warning performance would additionally require an issuance rule and an explicit cost or loss function.

The threshold dependence also limits where a net score gain can be expected. The changes to the two Brier contributions were offsetting at the monthly 99th percentile. Above the historical record, the pointwise paired intervals favoured STAC in this sample, but the absolute differences decreased as exceedances became rarer. Moreover, matching the overall probability mass becomes harder to establish at the deepest threshold because the observed frequency is estimated from few exceedances, some of which verify on the same grid-cell date. The residual overprediction at the record threshold and in the shortest-tail locations illustrates why an occurrence ratio near one for the domain is an incomplete description of reliability.

\subsection{Choice of calibration target and longer lead times}

The rolling isotonic correction is a competitive baseline for users interested in one fixed threshold. Its record Brier score difference from STAC had an interval including zero. For applications that need probabilities across a range of temperatures, a forecast distribution serves a different purpose because its exceedance probabilities remain ordered as the threshold changes. The present case study tests a threshold-specific map alongside distributional post-processors. Non-parametric distributional approaches also merit comparison. For example, Biegert, Allen, Alber and Lerch (2026) used isotonic distributional regression to study the information about extremes in deterministic forecasts. A benchmark of such approaches for this application would require fitting the probability distributions on aligned training data and comparing their archived forecasts on the same cases.

The persistence of the WeatherNext 2 discrimination advantage at days 8--10 matters for forecast interpretation. A model can assign an approximately correct total probability across land cases while ranking the cases less effectively. Area under the receiver operating characteristic curve and the in-sample CORP score after isotonic recalibration both favour WeatherNext 2 with EMOS at longer leads in these comparisons. A monotone probability remapping can preserve rank order or introduce ties. Improving discrimination can require new predictors, a different backbone or a post-processor that changes the ranking. The tested systems differ in several of these respects. The comparison therefore supports a system-level conclusion about the observed long-lead gap.

\subsection{Predictor evidence and scope}

Removing the tail-weighted loss increased aggregate overprediction at longer leads, whereas the changes in Brier score remained uncertain. Removing the combined land, surface and preceding-window inputs increased short- to medium-range Brier loss, with an unresolved change in the occurrence ratio. This distinction argues for evaluating model components against the intended property of the forecast. It also limits a physical interpretation of the soil-moisture strata. For example, a spatial relationship between the historical tail gap and frequency bias is compatible with differences in climatology, seasonality and forecast errors across regions. The grouped intervention and exploratory single-seed checks do not provide isolated estimates of the effects of soil moisture, spatial receptive field or the output distribution.

Our principal results apply to a common evaluation window in 2022, on the coarse verification grid and for a daily maximum formed from four synoptic samples. The thresholds refer to grid-cell climatology over a fixed historical period. They describe different events from local-hourly station records or current records that are updated after each exceedance. The initialisation source comparison retains ERA5 as the reference and tests transfer within the available Pangu archives. The 2020 comparison provides a consistency check across a different set of initialisations; because some preprocessing statistics span the evaluation year and a few targets fall in early 2021, it is not a fully independent test. The results are also conditional on access to archived ERA5 values at the stated nominal valid-time cutoffs. Actual dissemination latency and performance against independent stations remain separate questions.

\section{Conclusions}

For the three EMOS configurations tested, improved overall daily maximum temperature forecasts coexisted with a larger probability mass above a fixed historical monthly record. The direction of the effect was clearest for WeatherNext 2, whose record frequency ratio shifted from underprediction to overprediction. Spatially conditioned post-processing of Pangu produced a smaller record Brier score than static Pangu EMOS or rolling WeatherNext 2 EMOS at days 1--7. The gain over WeatherNext 2 came from reduced probability loss on non-exceedances, with higher loss on exceedances. An online correction fitted only to the record threshold achieved a similar record score within the uncertainty of the paired comparison, and WeatherNext 2 with EMOS retained better discrimination at days 8--10. For evaluations of historical heat-record forecasts, reporting probability bias, outcome-specific score contributions and discrimination together makes these trade-offs visible.

\section*{Declarations}

\noindent\textbf{Funding.} This research received no specific grant from any funding agency in the public, commercial, or not-for-profit sectors.

\noindent\textbf{Competing interests.} The authors declare no competing interests.

\noindent\textbf{Author contributions.} Y.W. conceived and led the study, developed the method, performed the analysis and wrote the manuscript. L.Z. and Y.H. contributed to data preparation and verification checks. All authors discussed the results, reviewed the manuscript and approved the final version.

\section*{Data availability}

All data used in this study are publicly available. ERA5 is available from the \href{https://doi.org/10.24381/cds.adbb2d47}{Copernicus Climate Data Store}. The forecast archives are distributed through \href{https://weatherbench2.readthedocs.io/en/latest/data-guide.html}{WeatherBench 2}:

\begin{itemize}
\item ERA5 \textperiodcentered{} \href{https://storage.googleapis.com/weatherbench2/datasets/era5/1959-2023_01_10-6h-240x121_equiangular_with_poles_conservative.zarr/.zmetadata}{six-hourly 1.5\ensuremath{{}^\circ} reanalysis}.
\item Pangu-Weather \textperiodcentered{} \href{https://storage.googleapis.com/weatherbench2/datasets/pangu/2018-2022_0012_240x121_equiangular_with_poles_conservative.zarr/.zmetadata}{ERA5-initialised forecasts} and \href{https://storage.googleapis.com/weatherbench2/datasets/pangu_hres_init/2022_0012_240x121_equiangular_with_poles_conservative.zarr/.zmetadata}{2022 HRES-initialised forecasts}.
\item IFS-ENS \textperiodcentered{} \href{https://storage.googleapis.com/weatherbench2/datasets/ifs_ens/2018-2022-240x121_equiangular_with_poles_conservative.zarr/.zmetadata}{2018--2022 ensemble forecasts}.
\item WeatherNext 2 \textperiodcentered{} \href{https://storage.googleapis.com/weatherbench2/datasets/fgn/2022-240x121_equiangular_with_poles_conservative.zarr/.zmetadata}{2022 ensemble forecasts}.
\end{itemize}

The analysis uses only these public archives. Code and trained model weights are not publicly available.

\section*{References}
\noindent Alet, F., Price, I., El-Kadi, A., Masters, D., Markou, S., Andersson, T.R., et al. (2025) Skillful joint probabilistic weather forecasting from marginals. \textit{arXiv} preprint. \href{https://doi.org/10.48550/arXiv.2506.10772}{DOI}

\noindent Allen, S., Ginsbourger, D. and Ziegel, J. (2023) Evaluating forecasts for high-impact events using transformed kernel scores. \textit{SIAM/ASA Journal on Uncertainty Quantification}, 11, 906--940. \href{https://doi.org/10.1137/22M1532184}{DOI}

\noindent Allen, S., Koh, J., Segers, J. and Ziegel, J. (2025) Tail calibration of probabilistic forecasts. \textit{Journal of the American Statistical Association}. \href{https://doi.org/10.1080/01621459.2025.2506194}{DOI}

\noindent Biegert, T., Allen, S., Alber, A. and Lerch, S. (2026) Towards fair comparisons of AI- and physics-based weather models for extreme events via the weighted potential CRPS. \textit{arXiv} preprint. \href{https://doi.org/10.48550/arXiv.2606.21170}{DOI}

\noindent Biegert, T., Allen, S., Zhang, Z. and Lerch, S. (2026) Can we trust evaluations of AI weather forecasts for extremes? How evaluation protocols shape model rankings. \textit{arXiv} preprint. \href{https://doi.org/10.48550/arXiv.2610.31237}{DOI}

\noindent Bi, K., Xie, L., Zhang, H., Chen, X., Gu, X. and Tian, Q. (2023) Accurate medium-range global weather forecasting with 3D neural networks. \textit{Nature}, 619, 533--538. \href{https://doi.org/10.1038/s41586-023-06185-3}{DOI}

\noindent Brier, G.W. (1950) Verification of forecasts expressed in terms of probability. \textit{Monthly Weather Review}, 78, 1--3. \href{https://doi.org/10.1175/1520-0493\%281950\%29078\%3C0001\%3AVOFEIT\%3E2.0.CO\%3B2}{DOI}

\noindent Dimitriadis, T., Gneiting, T. and Jordan, A.I. (2021) Stable reliability diagrams for probabilistic classifiers. \textit{Proceedings of the National Academy of Sciences}, 118, e2016191118. \href{https://doi.org/10.1073/pnas.2016191118}{DOI}

\noindent Ferro, C.A.T. (2014) Fair scores for ensemble forecasts. \textit{Quarterly Journal of the Royal Meteorological Society}, 140, 1917--1923. \href{https://doi.org/10.1002/qj.2270}{DOI}

\noindent Gneiting, T., Raftery, A.E., Westveld, A.H. III and Goldman, T. (2005) Calibrated probabilistic forecasting using ensemble model output statistics and minimum CRPS estimation. \textit{Monthly Weather Review}, 133, 1098--1118. \href{https://doi.org/10.1175/MWR2904.1}{DOI}

\noindent Hersbach, H., Bell, B., Berrisford, P., Hirahara, S., Hor\'{a}nyi, A., Mu\~{n}oz-Sabater, J., et al. (2020) The ERA5 global reanalysis. \textit{Quarterly Journal of the Royal Meteorological Society}, 146, 1999--2049. \href{https://doi.org/10.1002/qj.3803}{DOI}

\noindent Lang, S., Alexe, M., Clare, M.C.A., Roberts, C., Adewoyin, R., Ben Bouall\`{e}gue, Z., et al. (2026) AIFS-CRPS: ensemble forecasting using a model trained with a loss function based on the continuous ranked probability score. \textit{npj Artificial Intelligence}, 2, 18. \href{https://doi.org/10.1038/s44387-026-00073-7}{DOI}

\noindent Lerch, S., Thorarinsdottir, T.L., Ravazzolo, F. and Gneiting, T. (2017) Forecaster's dilemma: extreme events and forecast evaluation. \textit{Statistical Science}, 32, 106--127. \href{https://doi.org/10.1214/16-STS588}{DOI}

\noindent Pic, R., Dombry, C., Naveau, P. and Taillardat, M. (2025) Distributional regression U-Nets for the postprocessing of precipitation ensemble forecasts. \textit{Artificial Intelligence for the Earth Systems}, 4. \href{https://doi.org/10.1175/AIES-D-24-0067.1}{DOI}

\noindent Rasp, S., Hoyer, S., Merose, A., Langmore, I., Battaglia, P., Russell, T., et al. (2024) WeatherBench 2: a benchmark for the next generation of data-driven global weather models. \textit{Journal of Advances in Modeling Earth Systems}, 16, e2023MS004019. \href{https://doi.org/10.1029/2023MS004019}{DOI}

\noindent Trotta, B., Johnson, R., de Burgh-Day, C., Hudson, D., Abellan, E., Canvin, J., et al. (2025) Statistical postprocessing yields accurate probabilistic forecasts from artificial intelligence weather models. \textit{Artificial Intelligence for the Earth Systems}, 4, e250037. \href{https://doi.org/10.1175/AIES-D-25-0037.1}{DOI}

\noindent Van Velthoven, B., Decancq, C., Miralles, D.G., Luca, S. and Mortier, T. (2026) Beyond the record: evaluating probabilistic AI weather forecasts deep in the tail. \textit{arXiv} preprint. \href{https://doi.org/10.48550/arXiv.2610.25545}{DOI}

\noindent Wessel, J.B., Spuler, F., Ferro, C.A.T., Kretschmer, M. and Sippel, S. (2026) Probabilistic AI models outperform numerical weather prediction models on record breaking heat extremes. \textit{arXiv} preprint. \href{https://doi.org/10.48550/arXiv.2610.20519}{DOI}

\noindent Zhang, Z., Fischer, E., Zscheischler, J. and Engelke, S. (2026) Physics-based models outperform AI weather forecasts of record-breaking extremes. \textit{Science Advances}, 12, eaec1433. \href{https://doi.org/10.1126/sciadv.aec1433}{DOI}

\end{document}